# Comparison of skyrmion phases between poly and single-crystal MnSi by composite magnetoelectric method

Peipei Lu[1,2], Haifeng Du[3], Le Wang[4], Hang Li[4], Wenhong Wang[4], Youguo Shi[4], Xueliang Wu[2], Young Sun[2,4] * and Yisheng Chai[2]*

[1]*College of Physics and Hebei Advanced Thin Films Laboratory, Hebei Normal University, Shijiazhuang 050024, Hebei, China*

[2]*Low Temperature Physics Laboratory, College of Physics, and Center of Quantum Materials and Devices, Chongqing University, Chongqing 401331, China*

[3]*High Magnetic Field Laboratory, Chinese Academy of Science, Hefei 230031, Anhui, China*

[4]*Beijing National Laboratory for Condensed Matter Physics, Institute of Physics, Chinese Academy of Sciences, Beijing 100190, China*

## Abstract

We have explored the skyrmion phases and phase diagram of poly and single-crystal MnSi by the measurements of the magnetoelectric coefficient $\alpha_E$ and ac magnetic susceptibility of the MnSi/PMN-PT composite. We found that the regular skyrmion lattice phase in single crystal sample has been averaged in the MnSi polycrystal due to random grain orientations which results in an extended skyrmion lattice-conical mixture phase down to 25 K. The magnitude of the out-of-phase component in $\alpha_E$ of the polycrystal, not single crystal, decreases gradually with decreasing frequency. With the changing of the driven ac field, it reveals a depinning threshold behavior in both samples. The depinning field is stronger in the polycrystal than that in single crystal and maybe responsible for the diminishing of dissipative behavior at lower frequency due to grain boundaries and defects. The composite magnetoelectric method provides a unique approach to probe topological phase dynamics.

*Corresponding authors: youngsun@cqu.edu.cn, yschai@cqu.edu.cn

It is proposed that the assembly of some quasiparticles in solids, e.g., skyrmions in magnetic and ferroelectric materials and vortices in type-II superconductors, could exhibit two-dimensional close-packed solid phase.[1-4] In vortices solid, the existence of impurities or static disorders in crystal can always affect their dynamical properties. A finite pinning force density $F_{pin}$ by external driven field can be very different between weak defects and grain boundaries.[1] Meanwhile, the magnetic ac susceptibility in this region can also show distinctively different frequency dependent behaviors.[1,5] Therefore, it is very important to study the effects of various pinning centers in such systems for applications.

In analogy with the vortex phases in type-II superconductors, a magnetic skyrmion lattice phase has been identified in MnSi by Neutron, Hall and ac magnetic susceptibility, *etc*.[4,6-11] The existence of regular lattice pattern down to microscopic level has also been shown by Lorentz TEM.[12] By controlled doping, the characteristic of the disordered skyrmion phase in doped single crystals have been reported.[13-16] The existence of grain boundaries in the polycrystal sample to its skyrmion solid phase, particularly to its dynamic behavior, is still unclear. Recently, the coexistence of skyrmion lattice (**Sk-lattice**) and skyrmion-conical (**Sk-C**) coexistence phases were revealed in a MnSi single crystal by employing a composite magnetoelectric (ME) technique on the MnSi/PMN-PT laminate.[17] The disappearance of the out-of-phase component of the ME coefficient α deep inside the skyrmion (**Sk**) phase for the MnSi single crystal reveals the stabilization of **Sk-lattice** via the strong Sk-Sk interaction and low density of pinning centers. Therefore, this method should be applicable to compare the skyrmion phases between single and polycrystalline MnSi with weak and strong pinning centers, respectively.

In this work, the composite ME technique was used to compare the skyrmion phases and their dynamic behaviors of a poly and a single-crystal MnSi. By measuring the ac magnetic susceptibility χ and the ac ME voltage coefficient α of both samples, we find that the **Sk-lattice** and **Sk-C** coexistence phases in single-crystal samples are averaged to an extended skyrmion lattice-conical mixture (extended **Sk-C**) phase because of the random grain orientations in MnSi polycrystal. Moreover, by investigating the driven field dependent magnetoelectric coefficients, grain boundaries in polycrystal lead to a stronger depinning threshold and a subsequently diminishing of dissipative behavior at lower frequency.

The polycrystalline MnSi (Poly-MnSi) sample was synthesized by electromagnetic induction melting in the stoichiometric composition in an argon atmosphere. The sample was cut into a rectangular shape with a typical dimension 2 × 1 × 0.5 mm$^3$. The single-crystal MnSi (Single-MnSi) sample was synthesized using the Ga self-flux method.[17] The measurements of the ME voltage coefficient α were carried out in a Cryogen-free Superconducting Magnet System (Oxford Instruments, Teslatron PT). The Poly- and Single-MnSi/0.7Pb(Mg$_{1/3}$Nb$_{2/3}$)O$_3$–0.3PbTiO$_3$ (PMN-PT) laminates were prepared by silver epoxy (Epo-Tek H20E, Epoxy Technology Inc.) deposited onto two faces of [001]-cut PMN-PT single crystal ($t$=0.2 mm, thickness of the PMN-PT), and the longitudinal ME voltage coefficient $α_{E33}$ = d$E_3$/d$H_3$ was measured, as shown in Fig. 1(a). They were pre-poled by a Keithley 6517B electrometer with 650 kV/m at room temperature. The ac magnetic fields $H_{ac}$ = 3 and 1 Oe for poly- and single-crystal samples, respectively, were generated by a Helmholtz coil connected to a Keithley 6221 ac current source, as shown in Fig. 1(b). The ac ME voltage $V_{ac}$ was measured by a lock-in amplifier (Stanford Research SR830) as a function of dc magnetic field $H_{dc}$ or temperature $T$.[18,19] The α (α =$α_x$ + i$α_y$, in-phase component $α_x$, out-of-phase component $α_y$) was calculated by α = $V_{ac}$/($H_{ac}t$).

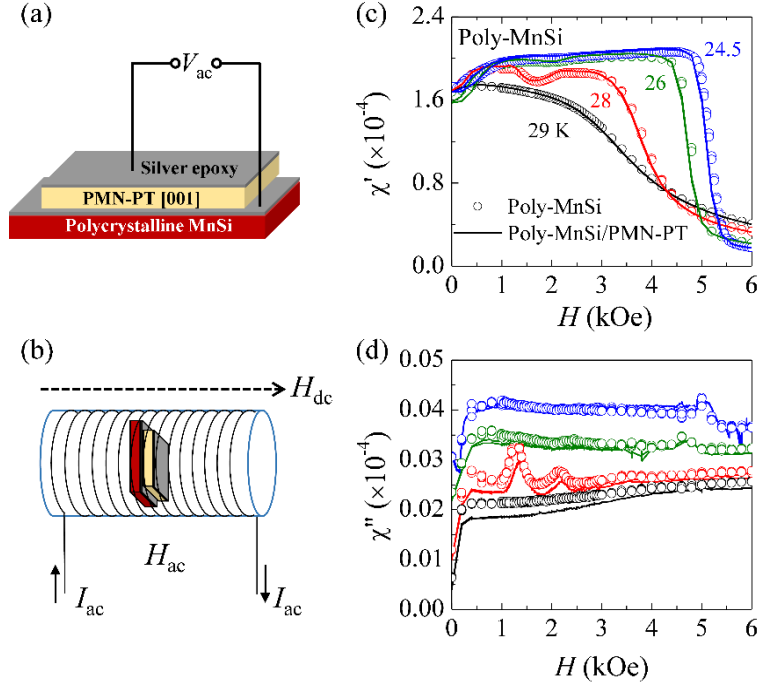

**FIG. 1.** (a) The structure of the Poly-MnSi/PMN-PT laminate. (b) The schematic measurement setup to supply ac magnetic field $H_{ac}$ and dc magnetic field $H_{dc}$. Magnetic field dependence of the ac magnetic susceptibility (c) χ' and (d) χ'' of Poly-MnSi with and without PMN-PT at selected temperatures under the driven frequency of 997 Hz.

The magnetic ac susceptibility (χ = χ'+iχ'') measurements were performed in a Magnetic Properties Measurement System (MPMS-XL, Quantum Design) with $H_{ac}$ = 3 and 1 Oe for poly and single-crystal samples, respectively. To check whether the skyrmion phase of polycrystalline MnSi is influenced by stress from PMN-PT layer in Poly-MnSi/PMN-PT laminate, the $H_{dc}$ dependent χ' and χ'' curves of Poly-MnSi/PMN-PT laminate and of Poly-MnSi itself at selected temperatures of 24.5, 26, 28 and 29 K were measured, as shown in Figs. 1(c) and 1(d). For χ'-H curves at selected temperatures, their magnitudes and behaviors of Poly-MnSi/PMN-PT and Poly-MnSi are almost the same. For χ''-H curves, the magnitudes are slightly different, but the main features and field dependent trends at each temperature are identical. To conclude, the influence of PMN-PT/epoxy on MnSi polycrystal is too weak to noticeably affect the skyrmion phases.

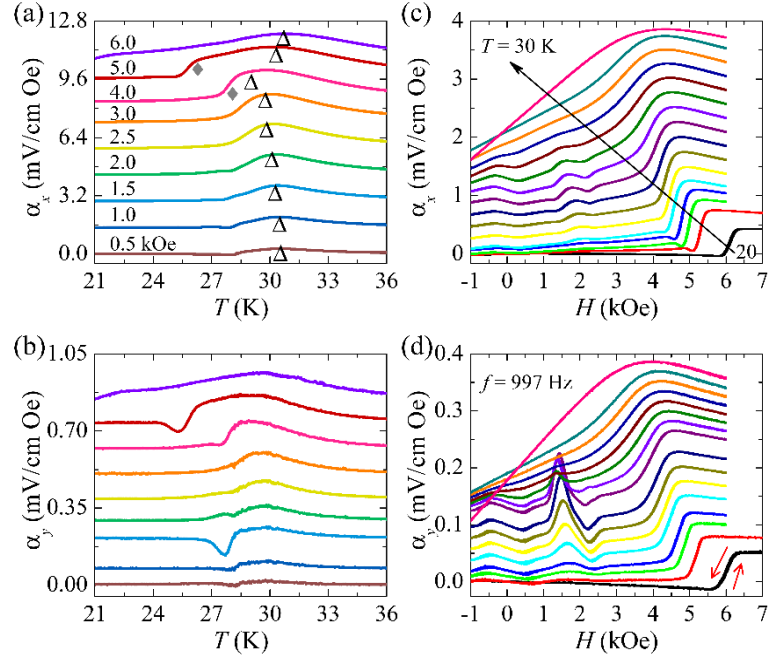

**FIG. 2.** The ac ME voltage coefficient components (a) $\alpha_x$ and (b) $\alpha_y$ of Poly-MnSi/PMN-PT laminate as a function of temperature under selected $H_{dc}$ at 997 Hz. Magnetic field dependence of (c) $\alpha_x$ and (d) $\alpha_y$ at selected temperatures of 20, 25, 26, 26.5, 27, 27.2, 27.4, 27.6, 27.8, 28, 28.2, 28.4, 28.6, 28.8, 29 and 30 K.

To investigate the skyrmion phases in polycrystalline MnSi, a MnSi sample with many grain boundaries as the source of disorders is prepared. We first measured the ME voltage coefficient α of Poly-MnSi/PMN-PT laminate as a function of temperature under selected $H_{dc}$ to roughly probe its magnetic phases and phase diagram, as shown in Figs. 2(a) and 2(b). The applied driven frequency is 997 Hz. For $\alpha_x$ under $H_{dc}$ from 4.0 to 6.0 kOe, there are broad peaks marked with open triangles, indicating the transition from ferromagnetic (**F**) to paramagnetic (**PM**) phase, in accordance with that of the MnSi single crystal.[17] This broad peak marks the transition from **PM** to a short-range correlation region (**Short-range**) below 4.0 kOe.[17,20] Under 4.0 and 5.0 kOe at 28 and 26.2 K, there is another broad bump signed as solid gray diamond at lower temperatures representing the transition between conical (**C**) and **F** phase. For $\alpha_y$, there are weak features similar to that of $\alpha_x$, indicating a small phase angle due to the measurement system. Figures 2(c) and 2(d) show the magnetic field-dependent $\alpha_x$ and $\alpha_y$ of the Poly-MnSi/PMN-PT laminate between 20 and 30 K, respectively. At 30 K, it enters into the **Short-range** phase with strong spin fluctuation below 4.0 kOe. Below $T_c$ = 29 K, very similar to that of Single-MnSi, there are clear features at every magnetic phase transition field.[17] In particular, the **Sk** phase of the polycrystalline MnSi exists between 25 and 29 K, marked by peak-dip features in both $\alpha_x$ and $\alpha_y$ curves, slightly wider than that reported in literatures.[4,10-12] Below 25 K, those peak-dip features eventually smear out.

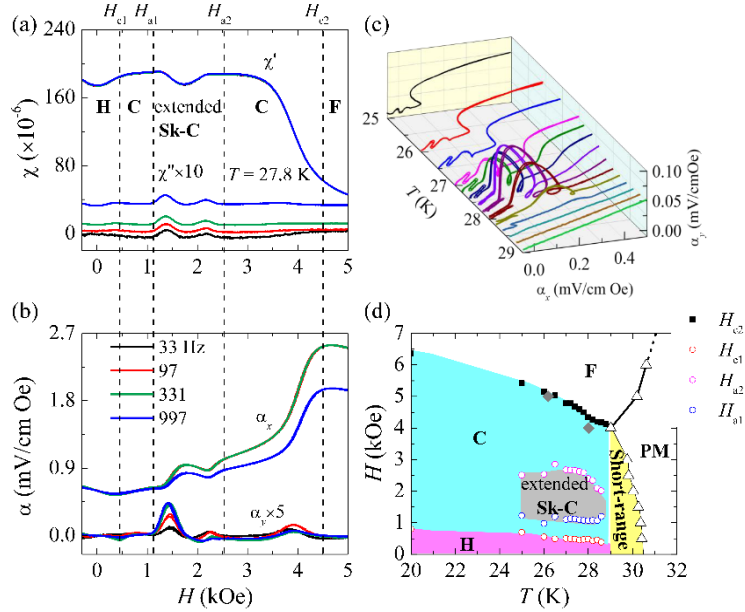

**FIG. 3.** (a) The magnetic field-dependent ac magnetic susceptibility ($\chi$) $\chi'$ and $\chi''$, (b) the ac ME voltage coefficient components of $\alpha_x$ and $\alpha_y$ of Poly-MnSi/PMN-PT at 27.8 K. (c) The Argand diagram of $\alpha_{E33}$ at selected temperatures. (d) The phase diagram of Poly-MnSi deduced from Fig. 2.

To further verify the peak-dip features marked skyrmion phase in the polycrystalline MnSi, we directly compare the $H_{dc}$ dependent ac magnetic susceptibility and $\alpha$ under four frequencies at 27.8 K. As shown in Fig. 3(a), $\chi'$ is dominant and frequency independent, indicating a very small phase lag in $\chi$ for all the magnetic phases. With decreasing frequency, the magnitude of $\chi''$ is almost invariant with some constant offsets. When $H_{dc}$ is swept from 4.5 to 2.6 kOe, there is a drastic increase in $\chi'$, indicating an **F** to **C** phase transition. The clear dip in $\chi'$ and two peaks in $\chi''$ from 2.6 to 1.1 kOe indicate the skyrmion phase (it is regarded as an extended **Sk-C** phase which will be explained later). The dip in $\chi'$ between 0.4 and 0 kOe points to a phase transition between **C** and the helical (**H**). All those features are similar to that of MnSi single crystal in literatures.[17,20-28] Figure 3(b) shows the $H_{dc}$ dependent $\alpha_x$ and $\alpha_y$ of the Poly-MnSi/PMN-PT laminate. In particular, the appearance of the peak-dip features at the intermediate field regime is consistent with the extended **Sk-C** phase in χ. Other features in $\alpha_x$ and $\alpha_y$ also match the phase boundaries found in magnetic susceptibility measurements.

To reveal whether the **Sk-lattice** can still exist under the influence of grain boundaries in the polycrystalline MnSi, we converted the field dependent complex $\alpha_{E33}$ data (Figs. 2(c) and 2(d)) into the Argand diagram, as shown in Fig. 3(c). In the single-crystal sample, the pure **Sk-lattice** phase has a non-dissipative nature that does not have any out-of-phase component in the Argand diagram.[17] In contrast, for the Argand diagram of polycrystalline MnSi, clear out-of-phase components appear in the entire skyrmion phases as well as around the transition regions between **C** and **F**. The polycrystalline MnSi is non-dissipative only in **H**, **C** and **Short-range** phases. The deduced *H-T* phase diagram of polycrystalline MnSi from the above measurements is summarized in Fig. 3(d). It nicely confirms the **H**, **C**, extended **Sk-C** and **F** phases in Fig. 3(a) and the **Short-range** phase from Figs. 2(a) and 2(b) in Poly-MnSi.

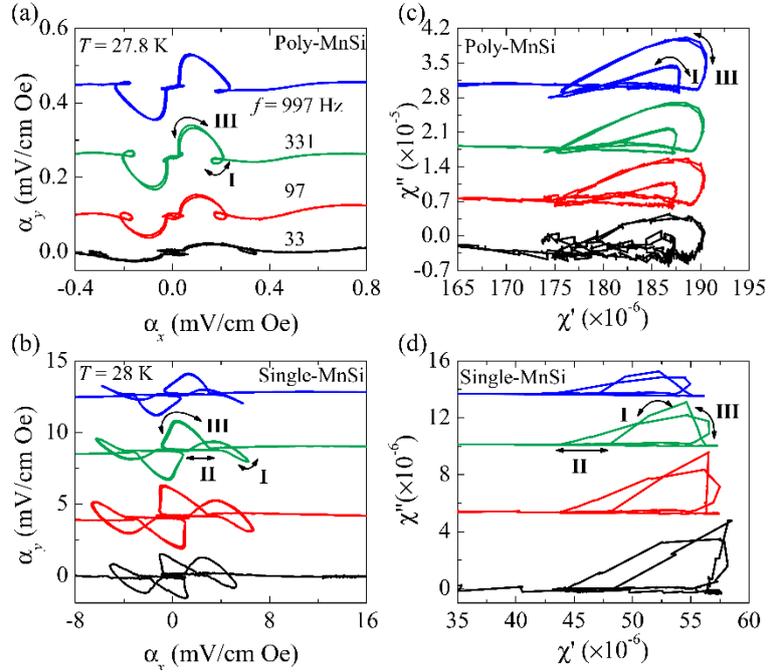

**FIG. 4.** The phase-corrected $\alpha_{E33}$ of (a) Poly-MnSi/PMN-PT at 27.8 K and (b) Single-MnSi/PMN-PT at 28 K, the χ of (c) Poly-MnSi/PMN-PT at 27.8 K and (d) Single-MnSi/PMN-PT at 28 K in the Argand plots. Data in (b) and (d) are from ref. 17.

To further confirm the extended **Sk-C** phase in the polycrystalline sample, α and χ of MnSi poly and single-crystal samples in the Argand plots at 27.8 K and 28 K, respectively, are compared under selected frequencies where the clear out-of-phase components in the Argand plot are most significant, as shown in Fig. 4. We have corrected the base angle (based on the high *T* data) in the Argand plot at each frequency by rotating the curve relative to the (0,0) point, as shown in Figs. 4(a) and 4(b). For data of Poly-MnSi at 27.8 K, an *H* scan can pass **H**, **C**, extended **Sk-C**, **C** and **F** phases successively. As we pointed out above, only the extended **Sk-C** phase allows the out-of-phase components in the Argand plot at each frequency. Particularly within the extended **Sk-C** phase, a larger positive out-of-phase component in section I connects to a negative out-of-phase component in section III without any intermediate non-dissipative section at each frequency. Similar two-section and three-section features of **Sk** phases in MnSi poly and single-crystal samples, respectively, are shown in magnetic susceptibility data in Figs. 4(c) and 4(d). It is very likely that, with respect to the applied magnetic field, the averaging of random grain orientations in MnSi polycrystal lead to an extended **Sk-C** phase.

We then look into the dynamical behaviors of Sk phases in poly and single-crystal samples. In detailed frequency dependent ME measurements, the magnitude of the out-of-phase components in $\alpha_{E33}$ of Poly-MnSi decreases gradually and becomes almost bump like at the lowest frequency of 33 Hz, indicating the diminishing of dissipative behavior at lower frequency. The strong weakening of out-of-phase component by lowering the driven frequency is consistent with the expected dynamic behavior of an extended **Sk-C** phase. These features are different from those of Single-MnSi/PMN-PT laminate at 28 K where the magnitude of out-of-phase components are almost invariant for different frequencies, as shown in Fig. 4(b).[17] Figure 4(c) shows the Argand plot of complex χ in Poly-MnSi/PMN-PT laminate at 27.8 K. Except the small offset along χ" axis, the magnitude of the out-of-phase components in extended **Sk-C** phase are almost identical at each frequency. The Argand plot of χ in Single-MnSi/PMN-PT laminate at 28 K slightly decreases with the increase of frequency, as shown in Fig. 4(d). The apparent discrepancy of frequency dependent behaviors between α and χ in polycrystalline sample may because that α is based on the magnetostrictive behavior of the material, showing obviously anisotropic. Thus, the ME coefficient are smeared out in a polycrystalline sample more than the magnetic susceptibility.

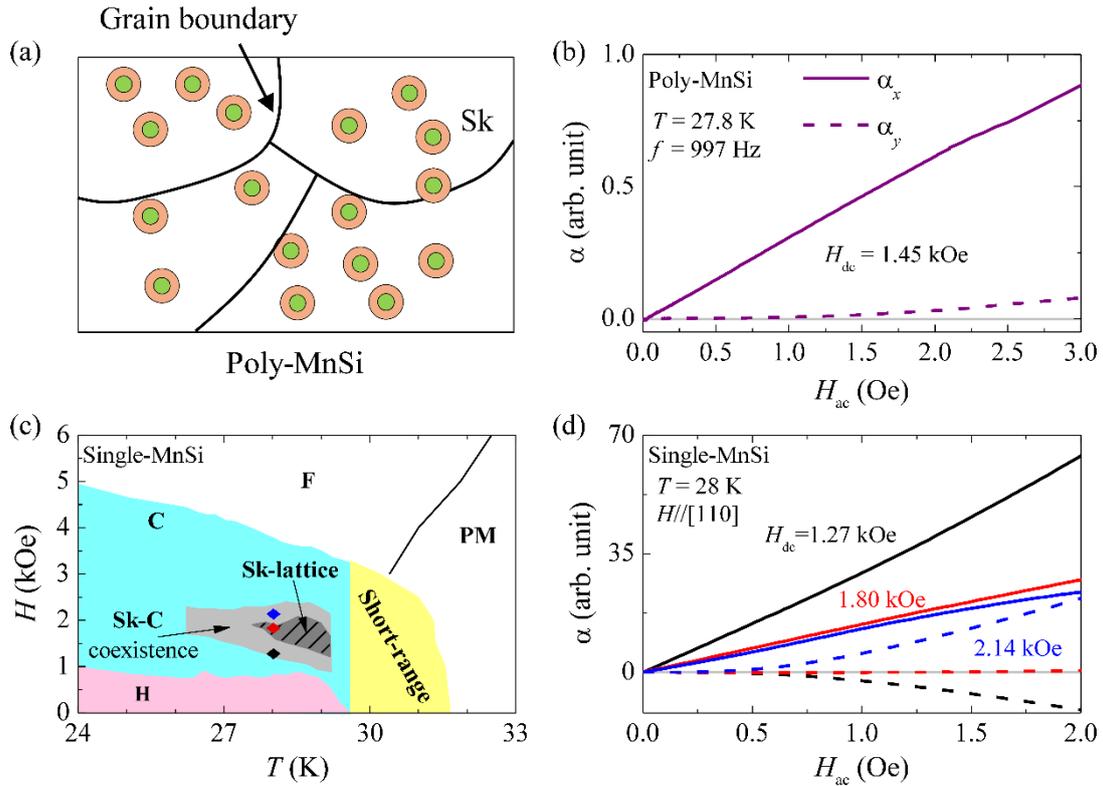

**FIG. 5.** (a) Schematic illustration of the extended **Sk-C** phase in Poly-MnSi. (b) The $H_{ac}$-dependent

$\alpha_x$ and $\alpha_y$ of Poly-MnSi/PMN-PT. (c) The magnetic phase diagram of a Single-MnSi sample in ref. 17. (d) $\alpha_x$ and $\alpha_y$ of Single-MnSi/PMN-PT as a function of $H_{ac}$ under various $H_{dc}$ of 1.27, 1.80 and 2.14 kOe. (c) is deduced from ref. 17.

We may postulate that the existence of grain boundaries is responsible for the distinctive dynamic behaviors in the polycrystal sample, as shown in Fig. 5(a). Thus, we compared the depinning behaviors between Sk phases in MnSi poly and single-crystal samples, as shown in Figs. 5(b)-(d). The ME voltage coefficients $\alpha$ as a function of ac driven field $H_{ac}$ are measured up to 3 and 2 Oe in poly and single-crystal samples, respectively, under selected $H_{dc}$. For Poly-MnSi, an $H_{dc}$ of 1.45 kOe at 27.8 K is chosen to have maximum $\alpha_y$ value in extended **Sk-C** phase. The $\alpha_y$ shows the expected depinning threshold behavior ($H_{ac}^{thr} \approx 1.2$ Oe) while the magnitude of $\alpha_x$ is linearly proportional to the driven magnetic field at smaller $H_{ac}$ value. The threshold value is broadened due to the grain boundaries. For drives $H_{ac} \ll H_{ac}^{thr}$, the Sk exhibits no depinning motion while the assembly of oscillating Sk particles around their pinning centers can still generate the elastic change of the sample and be reflected in $\alpha_x$. For $H_{ac} \sim H_{ac}^{thr}$, depinning occurs on some weakly interacted or weakly pinned Sks first, permitting the Sks to change neighbors over time, as shown in Fig. 5(a). For comparison, the phase diagram of MnSi single crystal is shown in Fig. 5(c).[17] At 28 K, three $H_{dc}$ = 1.27, 1.8, 2.14 kOe are chosen to locate in **Sk-C** coexistence, **Sk-lattice** and **Sk-C** coexistence phases, respectively. Clear depinning threshold behaviors ($H_{ac}^{thr} \approx 0.5$ and 0.4 Oe in I and III sections, respectively) are found while $\alpha_y$ is zero up to the $H_{ac}$ = 2 Oe for section II. The smaller depinning threshold field of the single-crystal sample marks a weak $F_{pin}$ while the non-threshold behavior in **Sk-lattice** is distinct from that of the extended **Sk-C** phase. In a word, the diminishing of dissipative behavior at lower frequency in polycrystal sample seems to be related to its stronger depinning threshold.

In summary, the Sk phase diagrams of a poly and a single-crystal MnSi were comprehensively investigated by the magnetic susceptibility and composite magnetoelectric technique. The skyrmion phases is basically unaffected by the stress from the PMN-PT layer. Furthermore, by comparing the α and χ data as a function of $H_{dc}$ between MnSi poly and single-crystal samples, the existence of extended **Sk-C** phase is indicated in polycrystalline MnSi. It is likely that grain boundaries are responsible for the unique dynamical behaviors of Sk phase in polycrystal sample.

## ACKNOWLEDGEMENTS

This work was supported by the National Natural Science Foundation of China (Grant Nos. 11974065, 52101221, 51725104), and Chongqing Research Program of Basic Research and Frontier Technology, China (Grant No. cstc2020jcyj-msxmX0263), Fundamental Research Funds for the Central Universities, China (2020CDJQY-A056, 2020CDJ-LHZZ-010, 2020CDJQY-Z006), the Natural Science Foundation of Hebei Province (Grant No. A2021205022), Science and Technology Project of Hebei Education Department (QN2021088), Hebei Normal University (Grant No. L2021B09). Y. S. Chai would like to thank the support from Beijing National Laboratory for Condensed Matter Physics. We would like to thank Miss G. W. Wang at Analytical and Testing Center of Chongqing University for her assistance.

## Conflict of Interest

The authors declare that they have no competing interests.